\documentclass[prb,preprintnumbers,amsmath,amssymb,unsortedaddress]{revtex4-1}

\usepackage{graphicx}
\usepackage{dcolumn}
\usepackage{bm}

\begin{document}


\title{Is the regime with shot noise suppression by a factor 1/3 achievable\\
in semiconductor devices with mesoscopic dimensions?}

\author{P. MARCONCINI, M. MACUCCI, D. LOGOTETA, M. TOTARO}
\affiliation
{Dipartimento di Ingegneria dell'Informazione, Universit\`a di
Pisa, Via Caruso 16,\\
Pisa, I-56122, Italy\\
p.marconcini@iet.unipi.it}

\begin{abstract}
We discuss the possibility of diffusive conduction and thus of
suppression of shot noise by a factor 1/3 in mesoscopic semiconductor devices
with two-dimensional and one-dimensional potential disorder, for which
existing experimental results do not provide a conclusive result. On the
basis of our numerical analysis, we conclude that it is quite difficult to
achieve diffusive transport over a reasonably wide parameter range,
unless the device dimensions are increased up to the macroscopic
scale. In addition, in the case of one-dimensional disorder,
some mechanism capable of mode-mixing has to be present in order to reach 
or even approach the diffusive regime.
\end{abstract}

\maketitle

\section{Introduction}

In nanoscale devices shot noise (i.e. the noise
due to the discreteness of the electron charge) is often suppressed
with respect to the value of power spectral density given by the
Schottky formula $S_I=2 e |I|$ (where $e$ is the elementary charge and
$I$ is the average current), which would be expected in the absence
of correlations between the carriers~\cite{schottky}. The ratio that 
quantifies this
suppression, due to the correlations introduced by Coulomb
interactions or (as in the cases discussed in this article) by Fermi
exclusion, is called Fano factor.

Particularly interesting is the case of diffusive conductors.
It is well known~\cite{datta} that, if the length $L$ of the device is much less
than the elastic mean free path $l$, the transport regime is ballistic
and shot noise is strongly suppressed,
while, if $L$ is greater than $N l$ (where $N$ is the number of
propagating modes) and the conductor is assumed to be phase coherent 
(as we will do in this work), transport is characterized by strong 
localization,
with the resistance increasing exponentially with the length of the
device and the Fano factor approaching unity.
Instead, in the intermediate regime in which the values of $l$
(determined by the disorder inside the device), $L$ and $N$
satisfy the inequalities $l \ll L \ll N l$, transport is diffusive. In this
regime the resistance increases linearly with length and, being the
distribution of the transmission eigenvalues $T$ bimodal and
proportional to $1/(T\sqrt{1-T})$, shot noise is suppressed by a
factor 1/3, due to the presence of a few open quantum channels with
nearly unit transmission probability~\cite{dejong1,nazarov}. This result, 
in the case of
devices with more than one-dimensional disorder, has been
theoretically obtained using quantum-mechanical~\cite{beenakker,altshuler} or 
semiclassical~\cite{nagaev} approaches, both of which include the effect of
Fermi exclusion. The same result has been obtained in the case of
one-dimensional disorder (i.e. a series of unevenly spaced tunnel
barriers) using a semiclassical model~\cite{dejong2}.

In this paper we will try to answer the question whether in semiconductor
devices with mesoscopic dimensions it is possible to reach the diffusive
transport regime and thus the corresponding 1/3 suppression of shot noise
over a wide parameter range.

From the experimental point of view, Henny {\em et al.}~\cite{henny} measured 
the shot noise suppression factor in a thin metallic wire, finding 
the theoretically predicted 1/3 reduction by asymptotically widening the 
contact reservoirs and thus reducing the reservoir heating which 
otherwise would increase the measured noise.

Instead, in the case of semiconductor devices, the existing experimental
data are not conclusive. Liefrink {\em et al.}~\cite{liefrink} reported,
for a wire obtained confining the two-dimensional electron gas of a
GaAs/AlGaAs heterostructure, shot noise reduction factors varying between
0.2 and 0.45, depending on the width of the wire. Instead 
Song {\em et al.}~\cite{song}
have measured the Fano factor for a semiconductor device with one-dimensional
disorder, in particular for a superlattice in which carriers have been
injected into the conduction band through optical excitations from the
valence band. Also in this case, for low applied fields, the measured
Fano factor has values that strongly differ from 1/3, varying
from one superlattice to another and in particular typically increasing
with barrier width (i.e. decreasing barrier transparency).

Prompted by these results, we have performed several numerical investigations
of the noise behavior of semiconductor structures characterized by 
2-dimensional (2D) and one-dimensional (1D) disorder, using different models, 
with varying degree of approximation.
In the case of 2D disorder, our conclusion is that in a semiconductor
device with mesoscopic dimensions it is very difficult to obtain a diffusive
behavior over a relatively large range of scatterer concentration or 
disorder strength:
in order to reach such a result in a definite way, we should  
increase the number of propagating modes up to several thousands and thus 
consider a macroscopic device.
On the other hand, in the case of 1D disorder we have found that the 
absence of coupling among the modes propagating in the structure makes it
inherently impossible to obtain a diffusive regime, and thus a 1/3 Fano
factor, unless some additional contribution (such as a magnetic field) 
is introduced to create mode mixing.  

\section{Two-dimensional disorder}

The structure we consider is a quantum wire obtained laterally confining,
by means of negatively biased gates located on the surface, the
two-dimensional electron gas (2DEG) of a GaAs/AlGaAs heterostructure.
The ionized donors located inside the n-doped AlGaAs layer (together
with other charged impurities present inside the heterostructure)
determine potential fluctuations at the 2DEG level.

In a previous paper by our group~\cite{bonci} a self-consistent calculation 
of the
average potential, combined with a semi-analytical formula for the effect of
ionized donors, was used to fit the conductance measurement
on a fabricated quantum wire and to numerically predict its noise behavior.
In that case it was found that the Fano factor did not stabilize at 1/3, 
but it rather crossed it for a single value of the gate bias voltage.

In order to gain a better understanding of the problem, we have now studied 
the noise behavior of the structure for a larger range
of parameters, for some choices of a model potential.
In detail, we have considered a 4.9~$\mu$m long and 8.4~$\mu$m wide
conductor with a hard-wall lateral confinement (since in our previous 
investigations the detailed shape of the confinement potential did not 
appear to play a significant role on the noise behavior).
We have considered that all of the dopants are located at a distance 
$D=40$~nm from the 2DEG and, for each considered dopant concentration (with 
a uniform random distribution), we 
have initially evaluated the effect, at the level of the 2DEG, by summing
up each individual contribution. The contribution of a single dopant 
has been evaluated with the
semi-analytical expression given by Stern and Howard~\cite{stern},
according to which a point charge $Ze$ located at a distance
$D$ from the 2DEG generates on the 2DEG, at a distance $r$ from its
orthogonal projection onto the 2DEG plane, a screened potential equal to
\begin{equation}
\phi (r)=\frac{Ze}{4 \pi \epsilon_0 \epsilon_r}
\int_0^\infty \frac{k}{k+s} J_0(k r) e^{-k D} dk
\end{equation}
where $\epsilon_0$ is the vacuum permittivity, $\epsilon_r$ is the
relative permittivity of the semiconductor, $J_0$ is the Bessel function
of order 0 and the screening constant is assumed equal to
$s=2 n_{\nu} (m^* e^2)/(4 \pi \epsilon_0 \epsilon_r \hbar^2 )$ (with
$n_{\nu}=1$ the considered subband degeneracy). 
In order to have a neutral system with a potential landscape symmetric
around zero, which simplifies the investigation and comparison of a  
large number of different cases, we have considered an 
artificial situation with an equal number of positive and negative charges.
Then, different disorder strengths for each dopant concentration have 
been obtained simply by multiplying the thus obtained potential profile 
by a scale factor $K$, thereby reducing the computational effort. 

\begin{figure}[th]
\centerline{\includegraphics[width=6cm]{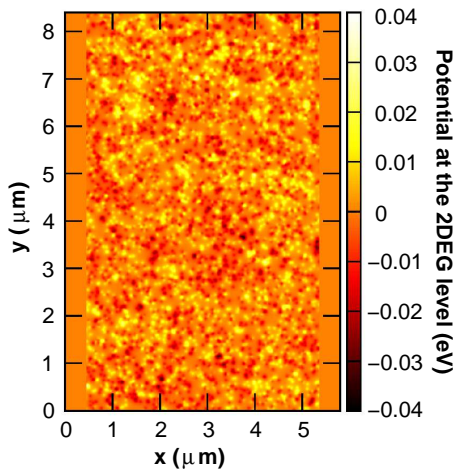}}
\vspace*{8pt}
\caption{Map of the potential obtained at the 2DEG level for a concentration
$N_D=1.1\times 10^{14}$~m$^{-2}$ of impurities located at a distance
$D=40$~nm from the 2DEG, multiplied by a scale factor $K=39$.}
\label{fig1}
\end{figure}

As an example, in Fig.~\ref{fig1} we show a
map of the potential obtained at the 2DEG level for a concentration
$N_D=1.1\times 10^{14}$~m$^{-2}$ of impurities located at a distance
$D=40$~nm from the 2DEG, multiplied by a disorder strength scale factor
$K=39$.

Once the potential at the 2DEG level has been obtained, the transmission 
matrix of the device has been evaluated using the recursive Green's function 
technique, with a representation over the transverse modes in the confined 
direction and in real space in the transport direction~\cite{sols,macucci1}. 
From the transmission 
matrix $t$, the conductance $G$, the shot noise power spectral density $S_I$ 
and the Fano factor $\eta$ have been obtained using the 
relations~\cite{buttiker}:
\begin{equation}
G=\frac{2\,e^2}{h}\,\sum_{i} w_i,\quad
S_I=4\,{e^3\over h}\,|V|\sum_i w_i(1-w_i),
\label{condnoi}
\end{equation}
\begin{equation}
\eta={S_I\over 2\,e\,|I|}=\frac{\sum_i w_i\,(1-w_i)}{\sum_i w_i},
\label{fano}
\end{equation}
where $V$ is the mean value of the externally applied voltage,
$h$ is Planck's constant and the $w_i$'s are the eigenvalues of the 
matrix $t^\dagger t$.

In these calculations we have separately averaged
the conductance and noise results (and thus the numerator and denominator
of Eq.~(\ref{fano})) over 41 energy values uniformly spaced in an energy range
of 80~$\mu$eV around $E_F=9$~meV.

\begin{figure}[th]
\centerline{\includegraphics[width=12.5cm]{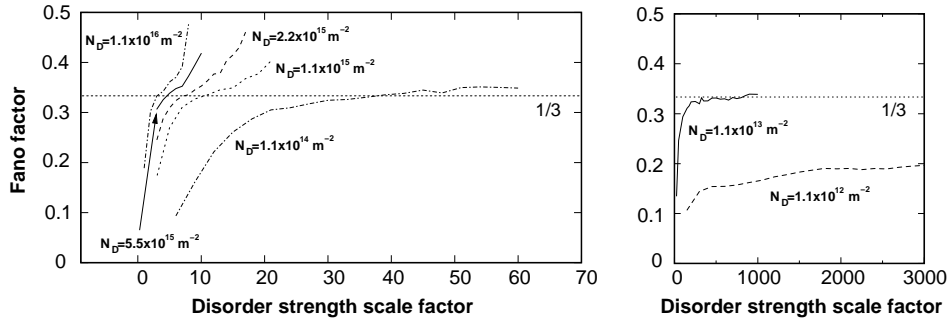}}
\vspace*{8pt}
\caption{Fano factor as a function of the disorder scale factor $K$ for 7
values of the dopant concentration $N_D$ and for $E_F=9$~meV.}
\label{fig2}
\end{figure}

In Fig.~\ref{fig2} we report the Fano factor that we have obtained for 7
values of the dopant concentration $N_D$, as a function of the
disorder strength scale factor $K$. We see that for elevated concentrations
the interval of disorder strength within which the curves approach the value
1/3 is very narrow, while it gets wider for low concentrations.
However, in this latter case the diffusive behavior is obtained only for
very large scale factors $K$. If we focus our attention on the potential
deriving, at the 2DEG level, from each single charged impurity and in
particular on the portion that more affects transport, i.e. that 
above the Fermi energy, we see that for these values of $K$ its spatial
extent is of the order of hundreds of nanometers. Since this extension is
unrealistic for semiconductor devices (while it could be reasonable for
metallic conductors, characterized by the presence of large grains), we 
conclude that in semiconductor nano-devices it is quite unlikely to 
obtain a 1/3 shot noise suppression factor within a reasonably large 
parameter range. Finally, for the lowest concentration 
($N_D=1.1 \times 10^{12}$~m$^{-2}$) the Fano factor
remains well below the 1/3 value and thus we have a substantially ballistic 
regime for all the considered disorder strengths.

\begin{figure}[th]
\centerline{\includegraphics[width=12.5cm]{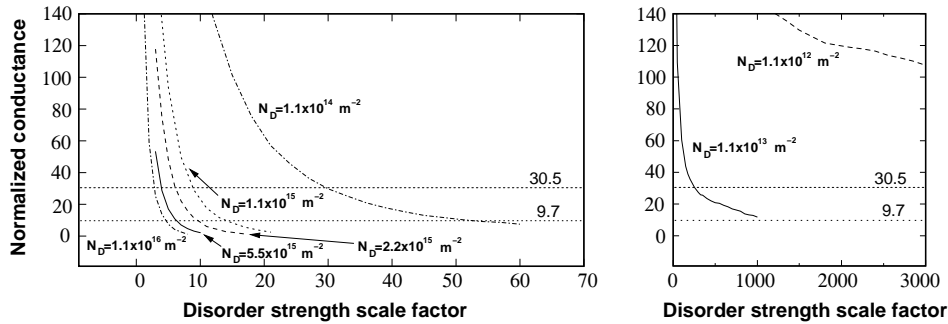}}
\vspace*{8pt}
\caption{Conductance (normalized with respect to the conductance quantum $G_0$)
as a function of the disorder scale factor $K$ for 7 values of the dopant
concentration $N_D$ and for $E_F=9$~meV.}
\label{fig3}
\end{figure}

In Fig.~\ref{fig3} we show the normalized conductance $G/G_0$ (where $G_0$
is the conductance quantum $2e^2/h$) as a function of the disorder strength
scale factor $K$ for the same impurity concentrations $N_D$.

The relationship between the conductance $G$, the mean free path $l$
and the device length $L$ is approximately given by $G/G_0=Nl/(L+l)$, with
$N$ the number of propagating modes~\cite{datta}. In this case, being $N=336$, the
condition for diffusive transport $l \ll L \ll N l$ is satisfied by a factor 
of 10 for both inequalities if $9.7<G/G_0<30.5$.

We again observe that the interval in which the condition
for diffusive transport is satisfied is very narrow for elevated 
concentrations, while it widens for low concentrations, for which, however, 
large disorder strengths are needed. Finally, in the case 
of $N_D=1.1 \times 10^{12}$~m$^{-2}$, the conductance never satisfies 
such a condition.

The comparison between Fig.~\ref{fig2} and Fig.~\ref{fig3} confirms that the
Fano factor assumes values near 1/3 in the same parameter range in which
the condition $l \ll L \ll N l$ is satisfied.
This range can clearly be enlarged increasing the number $N$ of propagating
modes, but this can be obtained only considering wider conductors, with
macroscopic dimensions,  thus outside the mesoscopic range we are interested 
in.

In the case of metallic conductors, due to the higher number of propagating
modes and to the possibility of extended scatterers, 
diffusive transport can actually be reached and experimentally
measured.

Finally, in order to compare these conclusions with some previously 
obtained results~\cite{macucci2},
we have performed some simulations using a more simplified model for the 
disordered potential. In detail, in Ref.~\cite{macucci2} a discussion on the 
exact conditions needed to obtain the diffusive transport regime was presented,
considering wires where the disordered potential was simulated with a
random distribution of square obstacles.

\begin{figure}[th]
\centerline{\includegraphics[width=7cm]{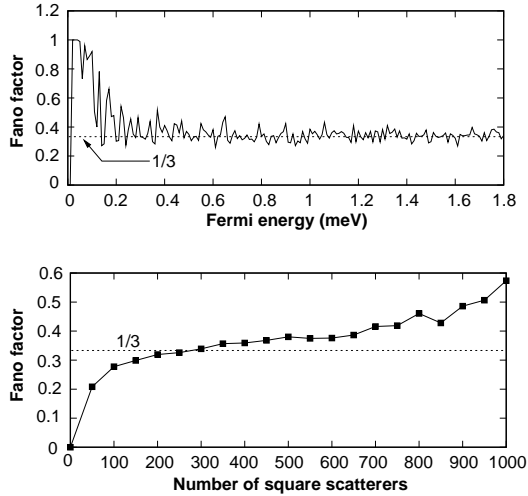}}
\vspace*{8pt}
\caption{
Upper panel: Fano factor as a function of the Fermi energy of the impinging
electrons, obtained in a $5~\mu$m wide and $8~\mu$m long wire,
containing 300 hard-wall 100~nm$\times$100~nm obstacles.\hfill\break
Lower panel: Fano factor as a function of the number of hard-wall square
scatterers in the wire.}
\label{fig4}
\end{figure}

In particular, we have repeated the calculation corresponding to the upper
curve of Fig.~6 of Ref.~\cite{macucci2}, obtaining the results reported in
the upper panel of Fig.~\ref{fig4}.
The simulation has been performed considering a wire with 
a width $W=5~\mu$m and a length $L=8~\mu$m, containing 300 hard-wall square 
obstacles, with a 100~nm edge. In the figure we show the Fano factor as 
a function of the Fermi energy of the impinging electrons. We see that, as soon 
as the Fermi energy has reached the value corresponding to a number 
of propagating modes for which the condition $l \ll L \ll N l$ is satisfied, 
the Fano factor reaches the value 1/3.

In order to verify whether the diffusive regime is preserved over a large 
range of scatterer concentrations, we have made some simulations varying 
the number of square scatterers inside the wire and computing the Fano factor 
around a value of the Fermi energy, $E_F=1.25$~meV, for which the diffusive 
regime has been reached in the case of 300 scatterers.
In particular, we have separately averaged the conductance and noise results
(and thus the numerator and denominator of Eq.~(\ref{fano}))
over a set of 41 uniformly spaced energy values in a range of 80~$\mu$eV
around 1.25~meV.
In the lower panel of Fig.~\ref{fig4} we show the behavior of the Fano factor
as a function of the number of square scatterers inside the conductor. We
see that also in this case the Fano factor does not settle around 1/3 for
a large range of scatterer numbers, but just crosses the 1/3 value in
correspondence of about 300 scatterers, which is indeed the situation for
which the conditions for diffusive regime have been investigated in
Ref.~\cite{macucci2}. Thus, this was a particular case, not representative 
of the general behavior.

\section{One-dimensional disorder}

The case of strictly one-dimensional disorder~\cite{marconcini1,marconcini2},
i.e. of unevenly spaced tunnel barriers located in an otherwise purely
ballistic device with any dimensionality, is a bit different from the cases
of 2D or 3D disorder, even though, according to a semiclassical study of
this structure~\cite{dejong2}, also in this case the shot noise suppression
factor should approach 1/3 as the number of cascaded barriers is let go 
to infinity.

A series of barriers can be defined, for example, in a heterostructure-based
device defining a series of gates on the surface of the device. Such gates,
when negatively biased, locally deplete the 2DEG, each generating a
tunnel barrier for the electrons traveling in the device.

We consider, for simplicity, idealized rectangular barriers, although we
have verified that the main results hold also for realistic 
barriers~\cite{totaro}. 
Since a wire with a series of such rectangular transversal barriers (each 
extending across the whole cross-section) can be seen as made up of a 
series of sections differing only for the value of their constant potential, 
the wave functions associated with the transverse modes are the same in all 
of the sections and the tunnel barriers do not introduce any mode-mixing.
Therefore the transport calculation can be subdivided into many purely 1D 
problems. Using a scattering matrix approach, in which the scattering matrix 
of each barrier and of each interbarrier region has a well-known analytical 
form, we have found the transmission $t_i$ for each mode $i$ through the 
device. The conductance, shot noise power spectral density and Fano factor 
can then be obtained using Eqs.~(\ref{condnoi})-(\ref{fano}), in which, due to 
the absence of mode-mixing, we have~\cite{lesovik} that $w_i=|t_i|^2$.

\begin{figure}[th]
\centerline{\includegraphics[width=6cm]{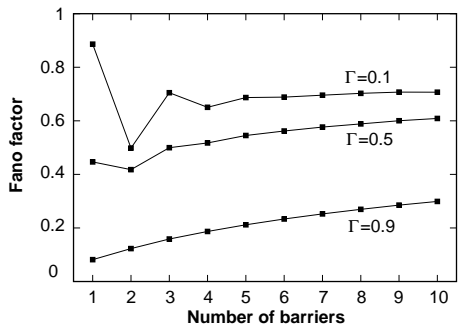}}
\vspace*{8pt}
\caption{Fano factor as a function of the number of 
barriers for 3 values of the barrier transparency $\Gamma$,
averaged over 50 different sets of interbarrier distances.}
\label{fig5}
\end{figure}

In Fig.~\ref{fig5} we report, for a series of identical barriers, the Fano
factor as a function of the number of the unevenly spaced barriers for 3 values
of the barrier transparency $\Gamma$. In detail, we have considered a 8~$\mu$m
wide structure and we have averaged our conduction and noise results over 500
energy values uniformly distributed in a range of 40~$\mu$eV around 9.03~meV.
We have considered 0.425~nm thick barriers, with heights equal to 0.8, 0.25
and 0.07~eV. Defining the barrier transparency $\Gamma$ as the squared
modulus of the transmission through each barrier, averaged over all the
propagating modes, for the 3 values of barrier height the transparency
$\Gamma$ is about equal to 0.1, 0.5 and 0.9,
respectively. In order to obtain a general behavior, we have averaged the
results over 50 different sets of interbarrier distances, which is 
equivalent to introducing dephasing with a simple phase randomization model
preserving localization effects~\cite{pala}.
These averages show that, contrary to what was expected from a semiclassical
analysis~\cite{dejong2}, no common asymptotic 1/3 value for the Fano factor
is reached increasing the number of barriers.

The reason is that in the absence of mode-mixing the overall transport
problem is just a collection of intrinsically one-dimensional problems;
therefore the localization length $L_l$ is equal to $l$ and thus it is 
impossible to satisfy the condition for diffusive transport $l \ll L \ll L_l$. 
Since semiclassical descriptions do not take into account the effect of 
phase coherence on transport and thus do not include localization, 
this effect can not be predicted using semiclassical arguments, but results 
only from a quantum-mechanical analysis.

In Ref.~\cite{totaro} we have also shown that 
the presence of a realistic amount of edge-roughness in the depletion gates
defining the barriers, introducing only a small degree of mode-mixing, does
not alter significantly the described results.

On the other hand, the amount of 2D disorder that we should add to the device
in order to create mode-mixing and reach the diffusive regime would be such
that it would lead to diffusive transport even in the absence of the barriers,
and therefore of the 1D disorder. 

Therefore, in order to reach the diffusive regime while preserving the 1D 
nature of the disorder, we have to introduce a different source of 
mode-mixing, for example a magnetic field threading the device. We have
shown elsewhere~\cite{marconcini1} that, especially in the case of barriers
with high transparency, the presence of a magnetic field
makes it possible to reach a 1/3 value for the Fano factor, by introducing 
mode-mixing and thus increasing the localization length with
respect to the mean free path.

\begin{figure}[t]
\centerline{\includegraphics[width=7cm]{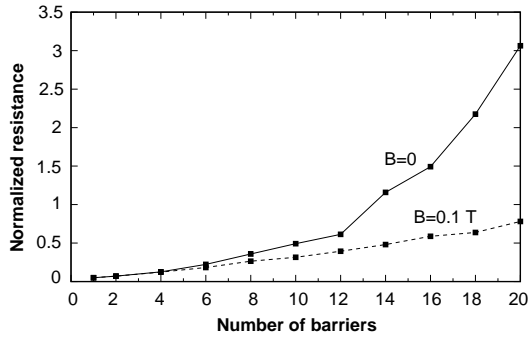}}
\vspace*{5pt}
\caption{Resistance (normalized with respect to the resistance quantum
$h/(2 e^2)$) as a function of the number of cascaded tunnel barriers in
a 1~$\mu$m wide conductor with 66~meV high and 1.56~nm thick barriers, with
$E_F=9.03$~meV; the exponential behavior has been obtained without the
magnetic field, while the nearly linear one has been found applying an
orthogonal magnetic field $B=0.1$~T.}
\label{fig6}
\end{figure}

Here we show the results obtained for a 1~$\mu$m wide conductor containing a
series of 66~meV high and 1.56~nm thick barriers, with an average transparency
at the considered Fermi energy (9.03~meV) $\Gamma=0.5$. The transport
calculation has been carried out using the recursive Green's function
technique, and adopting, for the representation of the vector potential, a
Landau gauge with nonzero component only along the transverse direction
\cite{marconcini3}. The conductance and noise results have been averaged over
a set of 25 energy values uniformly distributed over a range of 40~$\mu$eV
around 9.03~meV. The final results have been averaged over 20 different sets
of interbarrier distances. We see in Fig.~\ref{fig6} that, while in the absence
of magnetic field we observe an exponential behavior of the resistance as a
function of the number of the barriers (characteristic of the strong
localization regime), applying an orthogonal magnetic field $B=0.1$~T the
resistance behavior becomes approximately linear, i.e. we approach the
diffusive regime.

However, in order to obtain a diffusive regime over a really wide range of 
parameters, the mode-mixing introduced by the magnetic field has to be 
combined with the presence of a large number of propagating modes, which
requires, also in this case, macroscopic dimensions.

\section{Conclusion}

We have investigated the possibility of diffusive transport,
and thus of suppression of shot noise down to 1/3, in mesoscopic semiconductor
devices with 2D and 1D disorder.

From our numerical simulations, in which different representations for the
potential disorder have been adopted, we have concluded that it should be
very difficult and rather uncommon to obtain 
fully diffusive transport in mesoscopic semiconductor devices, 
due to the insufficient number of propagating modes.
In addition, in the case of 1D disorder the absence of mode-mixing 
makes it theoretically impossible to reach the diffusive regime,
unless a source of mode-mixing, such as a magnetic field, is present.

Our conclusions seem to be supported by existing experimental results, which
in the case of mesoscopic semiconductor devices have not shown a clear 1/3
suppression of shot noise.

\end{document}